\documentclass{ws-ijmpa}
\usepackage[super,compress]{cite}
\usepackage{graphicx}
\usepackage{amsmath}

\begin{document}
\markboth{K. A. Bronnikov \&\ V. G. Krechet}
	{Rotating cylindrical wormholes and energy conditions}

%
\catchline{}{}{}{}{}
%

\title{Rotating cylindrical wormholes and energy conditions}

\author{K. A. Bronnikov}

\address{Center for Gravitation and Fundam. Metrology, VNIIMS,\\
   	Ozyornaya St. 46, Moscow 119361, Russia;\\
	Institute of Gravitation and Cosmology, PFUR,\\
	Miklukho-Maklaya St. 6, Moscow 117198, Russia;\\
	National Research Nuclear University ``MEPhI''
	(Moscow Engineering Physics Institute),\\
	Kashirskoe sh. 31, Moscow 115409, Russia\\
	kb20@yandex.ru
	}

\author{V. G. Krechet}

\address{Moscow State Technological University ``Stankin'',\\
         Vadkovsky per. 3A, Moscow 127055, Russia}

\maketitle


\begin{abstract}
  We seek wormholes among rotating cylindrically symmetric configurations in
  general relativity. Exact wormhole solutions are presented with such
  sources of gravity as a massless scalar field, a cosmological constant,
  and a scalar field with an exponential potential. However, none of these
  solutions are asymptotically flat, which excludes the existence of
  wormhole entrances as local objects in our Universe. To overcome this
  difficulty, we try to build configurations with flat asymptotic regions
  using the cut-and-paste procedure: on both sides of the throat, a wormhole
  solution is matched to a properly chosen region of flat space-time at some
  surfaces $\Sigma_-$ and $\Sigma_+$. It is shown, however, that if the
  source of gravity in the throat region is a scalar field with an arbitrary
  potential, then one or both thin shells appearing on $\Sigma_-$ and
  $\Sigma_+$ inevitably violate the null energy condition.  Thus, although
  rotating wormhole solutions are easily found without exotic matter, such
  matter is still necessary for obtaining asymptotic flatness.

\keywords{General relativity, wormholes, rotation, cylindrical symmetry}
\end{abstract}

\ccode{PACS numbers: 04.20.Gz, 04.20.Jb, 04.40.Nr}

\def\GR{general relativity}
\def\cy{cylindrical}
\def\cyl{cylindrically symmetric}

\def\wh{wormhole}
\def\whs{wormholes}
\def\asflat{asymptotically flat}

\def\Jl#1#2{{\it #1} {\bf #2},\ }
\def\CQG#1 {\Jl{Class. Quantum Grav.}{#1}}
\def\GC#1 {\Jl{Grav. Cosmol.}{#1}}
\def\GRG#1 {\Jl{Gen. Rel. Grav.}{#1}}
\def\PRD#1 {\Jl{Phys. Rev. D}{#1}}
\def\PRL#1 {\Jl{Phys. Rev. Lett.}{#1}}

\def\cm{\hspace*{1cm}}
\def\inch{\hspace*{1in}}

\def\d{\partial}
\def\beq{\begin{equation}}
\def\eeq{\end{equation}}
\def\bear{\begin{eqnarray}}
\def\bearr{\begin{eqnarray} &&}
\def\ear{\end{eqnarray}}
\def\earn{\nonumber \end{eqnarray}}
\def\nn{\nonumber\\ {}}
\def\nnv{\nonumber\\[5pt] {}}
\def\nnn{\nonumber\\ \lal }
\def\yy{\\[5pt] {}}

\def\Half{{\dfrac{1}{2}}}
\def\half{{\tfrac{1}{2}}}
\def\sign{\mathop{\rm sign}\nolimits}
\def\diag{\mathop{\rm diag}\nolimits}
\def\const{{\rm const}}
\def\eps{\varepsilon}

\def\R{{\mathbb R}}
\def\kappa{\varkappa}
\def\e{{\rm e}}

\def\mn{_{\mu\nu}}
\def\MN{^{\mu\nu}}
\def\mN{_\mu^\nu}
\def\nM{_\nu^\mu}

\section{Introduction}

  Wormholes are hypothetic ``bridges'' or ``tunnels'' which connect
  different universes or different large or infinite regions of the same
  space-time. They are a subject of active discussion since their possible
  existence can lead to physical effects of great interest, such as
  realizable time machines or shortcuts between distant parts of the
  Universe, in particular, across black hole horizons\cite{thorne, viss-book,
  ws_book}. Unusual observable effects can be predicted under the assumption
  that \whs\ can exist on astrophysical scales of times and
  distances\cite{sha, astro, kir-sa1}.

  As is well known, the existence of a static \wh\ geometry in the framework
  of \GR\ requires the presence of ``exotic'', or phantom matter, that is,
  matter violating the weak and null energy condition (WEC and NEC), at
  least in a certain neighborhood of the throat
  \cite{thorne,viss-book,HV97,ws_book}, the narrowest place in a \wh. This
  conclusion, however, rests on the assumption that the throat is a compact
  2D surface, having a finite (minimum) area\cite{HV97}. In other words, a
  \wh\ entrance looks from outside as a local object like a star or a black
  hole.

  Since macroscopic exotic matter has not been observed in laboratory or
  in the Universe (except for the possible phantom dark energy), it is
  natural to try to obtain phantom-free (i.e., without matter violating the
  NEC and WEC) \whs\ by abandoning some of the assumptions of the
  Hochberg-Visser (HV) theorem\cite{HV97}. One of the simplest ways is to
  consider \cy\ symmetry thus rejecting the compact nature of the throats.
  Then, instead of starlike objects, we deal with objects like cosmic
  strings, infinitely extended along a certain direction. One can also
  consider nonstatic, rotating configurations, which can again be done in
  the framework of \cy\ symmetry.

  Examples of phantom-free \whs\ are known in some extensions of \GR, such
  as the Einstein-Cartan theory\cite{BGal15}, Einstein-Gauss-Bonnet
  gravity\cite{GBo}, brane worlds\cite{BKim03} and others. Nevertheless, it
  is highly desirable to further explore such an opportunity in \GR\ as a
  theory with the best experimental status, quite well describing the reality
  on the macroscopic scale.

  Cylindrical \whs\ with and without rotation in \GR\ were discussed, in
  particular, in Refs.~\citen{BLem09,BLem13} (see also references therein).
  It was shown there that phantom-free \cy\ \wh\ solutions to the Einstein
  equations are rather easily obtained, and there are numerous examples of
  such solutions, but none of them are \asflat.\footnote
	{Or at least \asflat\ up to an angular deficit appearing in
	cosmic string configurations. In what follows, for brevity, we will
	only mention asymptotic flatness although the same can be said about
	such cosmic-string asymptotics.}
  Meanwhile, asymptotic flatness is necessary for considering a wormhole
  entrance as a local (though extended in one direction) object in our
  Universe. To overcome this difficulty, we tried in Ref.~\refcite{BLem13}
  to build \wh\ configurations with flat asymptotic regions on both sides of
  the throat by matching a \wh\ solution to a properly chosen region of
  flat space-time at some surfaces $\Sigma_-$ and $\Sigma_+$. It was shown,
  however, that for \wh\ solutions with a massless scalar field such a
  procedure does not solve the problem since one or both thin shells
  appearing on $\Sigma_-$ and $\Sigma_+$ inevitably violate the NEC.

  The present paper generalizes Ref.~\refcite{BLem13} in two respects: (i)
  we present new exact \wh\ solutions with a scalar field as a source of
  gravity, namely, for a massless scalar in the presence of a cosmological
  constant and for a self-interacting field with an exponential potential,
  and (ii) we prove a no-go theorem saying that the above-mentioned
  cut-and-paste procedure does not lead to \asflat\ phantom-free \whs\ if
  the source of gravity in the \wh\ solution is a minimally coupled scalar
  field with an arbitrary potential.

\section{Basic Equations}

  A stationary \cyl\ metric with rotation can be written as
\bearr                                                    \label{ds-rot}
       ds^2 = \e^{2\gamma(u)}[dt - E(u)\e^{-2\gamma(u)}\, d\varphi]^2
\nn && \cm
       - \e^{2\alpha(u)}du^2 - \e^{2\mu(u)}dz^2 - \e^{2\beta(u)}d\varphi^2,
\ear
  where the second line gives the three-dimensional line element,
  $u$ is any admissible radial coordinate, $z\in \R$ and $\varphi \in
  [0,\ 2\pi)$ are the longitudinal and angular ones, respectively.
  There are two reasonable definitions of a \cy\ \wh\ throat:\cite{BLem09}
  (i) as a regular minimum of the circular radius $r(u) = \e^{\beta(u)}$
  (to be called an $r$-throat) and (ii) as a regular minimum of the area
  function $a(u) = \e^{\mu+\beta}$ (to be called an $a$-throat).

  A new feature of (\ref{ds-rot}) as compared to the static \cy\ metric
  (the same metric with $E \equiv 0$) is the emergence of a vortex
  gravitational field described as a 4-curl of the tetrad $e^\mu_a$: its
  kinematic characteristic is the angular velocity of tetrad rotation
  \cite{kr2}
\beq                                                      \label{def-o}
      \omega^\mu = \half \eps^{\mu\nu\rho\sigma} e_{m\nu} e^m_{\rho;\sigma},
\eeq
  where the Latin letters $m, n, \ldots$ stand for Lorentz indices. In our
  case, with an arbitrary radial coordinate $u$, the vortex $\omega
  = \sqrt{\omega_\alpha \omega^\alpha}$ is \cite{BLem13, kr2, kr4}
\beq                                                  \label{om}
     \omega = \half (E\e^{-2\gamma})' \e^{\gamma-\beta-\alpha}.
\eeq

  Furthermore, the off-diagonal component of the Ricci tensor $R_0^3$
  in the gauge $\alpha=\mu$ is given by
\beq                                                   \label{R_30}
	\sqrt{-g} R^3_0 = -(\omega \e^{2\gamma+\mu})',
	\ \ \ g := \det (g\mn).
\eeq
  Assuming that our rotating reference frame is comoving to the matter
  source of gravity, that is, the azimuthal flow $T^3_0 =0$, we find
  from $R^3_0 =0$ that
\beq       	      					\label{omega}
	\omega = \omega_0 \e^{-\mu-2\gamma}, \cm \omega_0 = \const,
\eeq
  and this relation is valid, by construction, in an arbitrary gauge.

  As a result,\cite{BLem13} the diagonal components of the Ricci tensor
  $R\mN$ can be written as the corresponding components ${}_s R\mN$ for
  the static metric (with $E=0$) plus the $\omega$-dependent addition
\beq
         {}_\omega R\mN = \omega^2 \diag (-2,\ 2,\ 0,\ 2),  \label{Ric-o}
\eeq
  The Einstein tensor $G\mN = R\mN - \half \delta\mN R$ splits
  in a similar manner, $G\mN = {}_s G\mN + {}_\omega G\mN$, where
\beq
	{}_\omega G\mN = \omega^2 \diag (-3,\ 1,\ -1,\ 1).  \label{Ein-o}
\eeq
  One can check that the tensors ${}_s G\mN$ and ${}_\omega G\mN$ (each
  separately) satisfy the ``conservation law''
  $\nabla_\alpha G^\alpha_\mu =0$ with respect to the static metric
  (with $E =0$).

  Then, according to the Einstein equations $G\mN = - \kappa T\mN$,
  the tensor ${}_\omega G\mN/\kappa$ behaves as an additional SET with
  quite exotic properties (thus, the effective energy density is
  $-3\omega^2/\kappa < 0$), acting in the auxiliary static metric
  (\ref{ds-rot}) with $E=0$. In its presence, it is rather easy to fulfil
  the throat existence conditions for both $r$- and $a$-throats, which is
  confirmed by some examples.\cite{kr4,BLem13}.

  In what follows we will give some new examples of \wh\ solutions using
  as a source of gravity a minimally coupled scalar field $\phi$ with
  a self-interaction potential $V(\phi)$. Its Lagrangian is
\beq							  \label{L_s}
	L_s = \frac 12 \eps \d_\alpha\phi \d^\alpha \phi -V(\phi),
\eeq
  where $\eps = +1$ corresponds to a normal scalar field and $\eps=-1$
  to a phantom one. Let us assume $\phi = \phi(u)$ and the comoving
  reference frame in the metric (\ref{ds-rot}), so that
  $T^3_0 =0$, so that the stress-energy tensor of $\phi$ is
\beq                                                     \label{SET1}
       T\mN (\phi) = \frac{\eps}{2}
	\e^{-2\alpha}\phi'^2 \diag(1,\ -1,\ 1,\ 1) + \delta\mN V(\phi).
\eeq
  Then, in the harmonic gauge\cite{kb79a,kb79b}
\beq
	\alpha = \beta + \gamma + \mu,               	\label{harm}
\eeq
  the Einstein-scalar equations take a particularly simple form, and
  their relevant combinations can be written as follows:
\bear                         	                        	\label{eq-s}
  	\eps \e^{-2\alpha} \phi'' &=& dV/d\phi,
\yy                             	                        \label{22}
	\e^{-2\alpha} \mu'' &=& \kappa V,
\yy                                     	                \label{03}
	\beta'' - \gamma'' &=& 4\omega_0^2 \e^{2\beta-2\gamma},
\yy                                             	        \label{023}
	2\mu'' - \beta'' - \gamma'' &=& 0,
\yy                                                             \label{int}
	\e^{-2\alpha}(\alpha'' - \alpha'{}^2 + \beta'{}^2
		+ \gamma'{}^2 + \mu'{}^2) &=&
		2 \omega^2 + \kappa \eps \e^{-2\alpha} \phi'^2 - \kappa V,
\ear
  where the prime denotes $d/du$, and (\ref{int}) is a first integral of the
  other four equations. Equations (\ref{023}) and (\ref{03}) are easily
  integrated giving
\bear                                                           \label{mu_}
	2\mu &=& \beta + \gamma + au, \ \ \ a = \const,
\yy
	\e^{\eta} &=& \frac{1}{2|\omega_0| s(k,u)}, \ \ \ \     \label{eta}
	    \eta := \beta - \gamma, \ \ \ \ k = \const,
\ear
  where two more integration constants have been excluded by choosing a
  scale along the $z$ axis and the origin of the $u$ coordinate, and the
  function $s(k, u)$ is defined as
\beq                                                          \label{def-s}
	s(k,u) =  \left\{
		\begin{array}{ll}
	k^{-1} \sinh ku, & \ \  k > 0,\ \ u \in \R_+;\\
		      u, & \ \ k=0, \ \ u \in \R_+; \\
	 k^{-1} \sin ku, & \ \ k<0, \ \ 0 < u < \pi/|k|.
		\end{array}\right.
\eeq

  As a result, the metric functions $\beta, \gamma, \mu$ are expressed in
  terms of $\eta(u)$ given by (\ref{eta}) and $\alpha(u)$ as follows:
\bear
	2\beta &=& \eta + \frac 13 (2\alpha - a u),          \label{beta}
\\
	2\gamma &=& - \eta + \frac 13 (2\alpha - a u),       \label{gamma}
\\
	2\mu &=& \frac 23 (\alpha + a u),                    \label{mu}
\ear
  while the remaining two unknowns $\phi(u)$ and $\alpha(u)$ obey the
  equations (\ref{eq-s}) and
\beq                                                          \label{eq-a}
	\e^{-2\alpha} \alpha'' = - 3\kappa V.
\eeq
  In addition, Eq.\,(\ref{int}) leads to their first integral
\beq
       \frac{\eps}{2}\kappa \phi'^2 = \kappa V \e^{2\alpha}  \label{int1}
		+ \frac{1}{3}\alpha'^2
		- \frac{1}{12} a^2 - \frac{1}{4} k^2 \sign k.
\eeq
  Lastly, the function $E(u)$ is easily found by integration from
  Eq.\,(\ref{om}) provided the other metric coefficients are known.

\section{Examples of Wormhole Solutions}

  One can make a general observation from the above equations without
  finding exact solutions: in the case $k < 0$, if a solution to
  Eqs.\,(\ref{eq-s}), (\ref{eq-a}), (\ref{int1}) is finite and regular on
  the segment $0 \leq u \leq \pi/|k|$ (or, equivalently, on any other
  half-wave of the function $\sin ku$), then the whole configuration is a
  \wh, with both $r$- and $a$-throats, and both $r(u)$ and $a(u)$ tend to
  infinity at the extremes $u =0$ and $u = \pi/|k|$, but these ends are
  singular due to $\e^\gamma \to 0$ and $\omega \to \infty$.

  We will confirm this observation with examples of exact solutions and also
  find that there can be \wh\ solutions with $k\geq 0$ as well.

\subsection{Vacuum and a massless scalar}

  To begin with, consider the solution without matter or with a massless
  scalar field corresponding to the Lagrangian (\ref{L_s}) with $V \equiv 0$.
  In this case, Eqs.\,(\ref{eq-s}) and (\ref{22}) read $\phi''=0$ and
  $\mu''=0$, which, combined with (\ref{mu_}), (\ref{eta}) and (\ref{int}),
  immediately lead to the complete solution\cite{kr4,BLem13,BS14}
\bear
       \e^{2\beta} &=& \frac{\e^{2hu}}{2|\omega_0| s(k,u)},\cm\ \qquad
       \e^{2\mu} = \e^{-2mu},
\nn                                                      \label{sol1}
       \e^{2\gamma} &=& 2|\omega_0| s(k,u) \e^{2hu},\cm\ \
       \e^{2\alpha} = \e^{(4h-2m)u},
\nn
       \omega &=& (\sign\omega_0) \frac{\e^{mu - 2hu}} {2s(k,u)},\cm
       			\phi = Cu
\nn
       E &=& \e^{2hu}[E_0 s(k,u) - s'(k,u)],
\ear
  where $\omega_0,\ E_0,\ h,\ k,\ m,\ C$ are integration constants obeying
  the relation
\beq                                                      \label{int-1}
	k^2 \sign k = 4(h^2 - 2hm) - 2\kappa C^2.
\eeq
  If the scalar charge $C$ is zero, it is a vacuum solution.

  In all branches of the solution, $r\to \infty$ and $\e^\gamma \to 0$ as
  $u\to 0$. In the same limit, the vortex $\omega \to \infty$, indicating a
  singularity. At the other end of the $u$ range, the situation is more
  diverse:

\begin{itemlist}
\item
     $k < 0$. A \wh\ geometry (with both kinds of throats) is described
     by all such solutions. At both ends, $\e^\beta \to \infty$
     and $\e^{\gamma}\to 0$, while $\e^{\beta + \gamma}$ and $\e^\mu$ remain
     finite, and $ \omega \sim \e^{-2\gamma} \to \infty$.
\item
     $k = 0$. At large $u$ we have $\e^{2\beta} \sim u^{-1}\e^{2hu}$ and
     $\e^{2\gamma} \sim u \e^{2hu}$, hence we have a \wh\ geometry with an
     $r$-throat if $h >0$ and with an $a$-throat if $h-m > 0$.
     In addition, $\e^\gamma \to \infty$ at large $u$.
\item
     $k > 0$. At large $u$, $\e^{2\beta} \sim \e^{(2h-k)u}$ and
     $\e^{2\gamma} \sim \e^{(2h+k)u}$, hence a \wh\ with an $r$-throat exists
     if $0 < k < 2h$; we also have $\e^\gamma \to \infty$ at large $u$.
     A \wh\ with an $a$-throat exists if $0 < k < 2(h-m)$.
\end{itemlist}

\subsection{A massless scalar and a cosmological constant}

  Consider the Lagrangian (\ref{L_s}) with $V = \Lambda/\kappa =\const$.
  Then Eq.\,(\ref{eq-s}) again takes the form $\phi'' = 0$, leading,
  as before, to $\phi = C u$, while (\ref{eq-a}) is a Liouville equation,
  like (\ref{023}), which is easily solved. Let us restrict ourselves to the
  case of larger interest $\Lambda > 0$, then Eq.\,(\ref{eq-a}) yields
\beq                                                        \label{a2}
	\e^{\alpha} = \frac{b}{\sqrt{3\Lambda} \cosh[b(u-u_0)]},
\eeq
  where $b > 0$ and $u_0 \in \R$ are integration constants. According
  to (\ref{beta})--(\ref{mu}),
\bear                                                           \label{sol2}
	 \e^{2\beta} &=& \frac{\e^{(-au + 2 \alpha)/3}}
	 {2|\omega_0| s(k, u)},
    \qquad
	 \e^{2\gamma} = 2|\omega_0| s(k, u)\,
	   \e^{(-au + 2\alpha)/3},
\nn
	 \e^{2\mu} &=& \e^{2(\alpha + au)/3},
    \qquad
	 \omega = \frac{(\sign \omega_0)}{s(k,u)}\,\e^{-\alpha}
\ear
  where $\e^\alpha$ is given by (\ref{a2}).
  The function $E(u)$ is found, as before, from (\ref{om}). The
  integration constants $k$, $a$, $b$, $C$ are connected by the following
  relation due to (\ref{int1}):
\beq
	 k^2 \sign k = \frac{4}{3}b^2 - \frac{1}{3}a^2 - 2\eps\kappa C^2.
\eeq

  This solution with $k < 0$ describes a \wh\ configuration according to the
  previous general observation. As to solutions with $k \geq 0$, we can look
  whether or not $r(u) \equiv \e^{\beta}$ and $a(u) \equiv \e^{\beta+\mu}$
  tend to infinity as $u \to \infty$ in order to check whether or not there
  are $r$- or $a$-throats, respectively. It is easy to verify that an
  $r$-throat is found if $3k + a + 2b < 0$, an $a$-throat if $3k -a + 4b <
  0$, and adding these two conditions, we conclude that a \wh\ with both
  kinds of throats exists only if $k + b < 0$, which is impossible since $k
  \geq 0$ by assumption and $b > 0$ according to (\ref{a2}). We conclude
  that this family of solutions describes \whs\ with both kinds of throats
  only in the case $k < 0$.

  For $\Lambda < 0$ the solutions are more diverse since Eq.\,(\ref{eq-a})
  leads to three branches similar to (\ref{eta}), but these solutions are
  beyond the scope of this paper.

\subsection{A scalar field with an exponential potential}

  Now consider the Lagrangian (\ref{L_s}) with the potential
\beq
	V(\phi) = V_0 \e^{2\lambda\phi}, \cm V_0 = \const > 0.
\eeq
  Two combinations of (\ref{eq-s}) and (\ref{eq-a}) are then easily
  integrable:
\bear                                                      \label{eq3-1}
	3\eps\kappa\phi'' + 2\lambda\alpha'' &=& 0,
\yy                                                        \label{eq3-2}
	(\alpha + \lambda\phi)'' &=&
		  (2\eps\lambda^2 - 3\kappa)V_0 \e^{2\alpha+2\lambda\phi}.
\ear
  Although solutions can readily be found for any values of the
  parameters involved, we will restrict ourselves to those with $\eps =
  +1$ (a non-phantom scalar) and $L^2 := V_0 (2\lambda^2 - 3\kappa) > 0$.
  The latter condition is justified since $\lambda$ is a length related to
  the field self-interaction and is likely to be of the order inherent to
  particle physics, so that $\lambda \gg l_p = \sqrt{G} =
  \sqrt{\kappa/(8\pi)}$ (the Planck length).

  Thus we find from (\ref{eq3-1}) and (\ref{eq3-2}):
\bear                                                      \label{sol3}
	3\kappa\phi + 2\lambda \alpha = C u,
\cm
	\e^{\alpha+\lambda\phi} = \frac{1}{L s(h, u - u_1)},
\ear
  where $C, h, u_1$ are integration constants and, as before, we suppress
  one more constant by choosing the zero point of $\phi$; the function
  $s(h, u-u_1)$ is defined by (\ref{def-s}) with proper substitutions.
  The integration constants are connected by the following relation due to
  (\ref{int1}):
\beq                                                        \label{int3}
	2C^2 - 12\kappa h^2 \sign h =
		(2\lambda^2-3\kappa) (a^2 + 3 k^2 \sign k).
\eeq
  For $\alpha(u)$ we obtain
\beq                                                          \label{a3}
	\e^{\alpha} = \e^{-\tfrac{\lambda C u}{2\lambda^2-3\kappa}}
	\Big[L\,s(h, u-u_1)\Big]^{-\tfrac{3\kappa}{2\lambda^2-3\kappa}}.
\eeq
  The other metric coefficients are now easily found using
  (\ref{beta})--(\ref{mu}), $\phi(u)$ is determined from the first equality
  in (\ref{sol3}), and then $\omega$ and $E$ from (\ref{omega}) and
  (\ref{om}), respectively.

  The solution behavior is rather diverse, depending on the interplay of
  zeros of the functions $s(k,u)$ and $s(h, u-u_1)$. We will not describe
  all variants in detail but only notice that there are \wh\ solutions
  with $k < 0$ according to the general description at the beginning of this
  section in all cases in which $s(h, u-u_1) > 0$ on the segment
  $0 \leq u \leq \pi/|k|$. There also exist some \wh\ solutions with
  $k \geq 0$, not to be considered here; a full description of this solution
  is postponed for future work.

\section{Asymptotic Flatness and Thin Shells: a No-Go Theorem}

  If we wish to describe \whs\ in our weakly curved Universe, potentially
  visible to distant observers like ourselves, it is necessary to suppose
  their flat (or string) asymptotic behaviors. However, it is hard to
  achieve when dealing with \cyl\ systems: indeed, even the Levi-Civita
  vacuum solution is \asflat\ only in the special case where the space-time
  is simply flat.

  A possible way out is to try to cut a non-\asflat\ \wh\ configuration at
  some regular cylinders $u=u_+$ and $u=u_-$ on different sides of the throat
  and to match it there to properly chosen flat space regions. The junction
  surfaces will then comprise thin shells with certain surface densities and
  pressures, and we should check whether or not they satisfy the
  standard energy conditions.

  To do so, one should take the flat-space metric in a rotating reference
  frame: in the Minkowski metric $ds^2 = dt^2 - dx^2 - dz^2 - x^2 d\varphi^2$,
  substituting $\varphi \to \varphi + \Omega t$, we obtain
\beq                                                          \label{ds_M}
      ds_{\rm M}^2 = dx^2 + dz^2 + x^2 (d\varphi + \Omega dt)^2 - dt^2,
\eeq
  $\Omega = \const$ being the angular velocity of the reference frame.
  The relevant quantities defined above are, in the notations of
  (\ref{ds-rot}),
\bear                                                   \label{M-param}
      g_{00} &=& \e^{2\gamma} =  1 - \Omega^2 x^2,
   \cm
      r^2 \equiv \e^{2\beta} = \frac{x^2}{1 - \Omega^2 x^2},
\nn
      E &=& \Omega x^2, \cm\ \omega = \frac{\Omega}{1 - \Omega^2 x^2}
\ear
  This metric is stationary and can be matched with an internal metric at
  $|x| <  1/|\Omega|$, inside the ``light cylinder'' on which the linear
  rotational velocity reaches that of light.

  Matching of two \cyl\ regions at a surface $\Sigma:\ u=u_0$ requires,
  above all, that this surface be the same as seen from both
  sides, therefore,
\beq                                                       \label{ju-1}
      [\beta] = 0, \qquad [\mu] = 0, \qquad
      [\gamma] = 0, \qquad [E] =0,
\eeq
  where, as usual, the square brackets denote discontinuities across the
  surface in question: for any $f(u)$, $[f] = f(u_0+0) - f(u_0 -0)$.
  One should note that in general the metrics on different sides of $\Sigma$
  may be written using different gauges (choices of the radial coordinate
  $u$), but it is unimportant since the quantities involved in all matching
  conditions used are insensitive to the choice of $u$.

  The next step is to determine the material content of the junction
  surface $\Sigma$ according to the Darmois-Israel
  formalism:\cite{israel-67,BKT-87}
  in our case of a timelike $\Sigma$, the surface stress-energy
  tensor $S_a^b$ is given by
\def\tK {{\tilde K}{}}
\beq                                                        \label{ju-2}
	S_a^b = - \frac{1}{\kappa} [\tK_a^b], \qquad
			\tK_a^b := K_a^b - \delta_a^b K,
\eeq
  where $K = K_a^a
$, $K_a^b$ is the extrinsic curvature of the surface
  $\Sigma$, and (since $\Sigma$ is $x^1 = \const$) the indices $a$ and $b$
  take the values $0,2,3$.

  Let us now assume that the internal region, containing both $r$- and
  $a$-throats, is described by a solution to the Einstein-scalar equations
  corresponding to the Lagrangian (\ref{L_s}) with a certain $V(\phi)$.
  The task is to choose the surfaces $\Sigma_\pm $, to perform matching
  and to calculate the surface densities and pressures.

  The matching conditions (\ref{ju-1}) on $\Sigma_\pm$, identifying the
  surfaces $x=x_\pm$ in Minkowski regions and $u=u_\pm$ in the internal
  region, are fulfilled by fixing the values of $x_\pm$ for given $u_\pm$,
  the scales along the $t$ and $z$ axes, and the values of $\Omega =
  \Omega_\pm$ in each Minkowski region. It is important that we should take
  $x_+ >0$ and $x_- < 0$ to adjust the directions of the normal vectors to
  $\Sigma_\pm$.

  Now, the question is whether the surface stress-energy tensors on
  $\Sigma_\pm$ can satisfy the WEC under some values of the system
  parameters. A criterion for that is the validity of the WEC which includes
  the requirements
\beq                                                             \label{WEC}
	\frac{S_{00}}{g_{00}} = \sigma \geq 0, \ \ \
		S_{ab}\xi^a \xi^b \geq 0,
\eeq
  where $\xi^a$ is any null vector ($\xi^a \xi_a =0$) on $\Sigma =
  \Sigma_\pm$, {\it i.e.,} the second inequality in (\ref{WEC}) comprises
  the NEC as part of the WEC. The conditions (\ref{WEC}) are equivalent to
\beq                                                            \label{WEC1}
      [\tK_{44}/g_{44}] \leq 0, \ \ \ [K_{ab}\xi^a \xi^b] \leq 0.
\eeq
  If we choose two null vectors on $\Sigma$ in the $z$ and $\varphi$
  directions as
\beq                                                        \label{xi_1,2}
	\xi^a_{(1)} = (\e^{-\gamma},\ \e^{-\mu},\ 0),
\ \ \ \
    \xi^a_{(2)} = (\e^{-\gamma}+ E\e^{-\beta-2\gamma},\ 0,\ \e^{-\beta}),
\eeq
  the conditions (\ref{WEC1}) read\cite{BLem13}
\beq                                                       \label{WEC2}
       [\e^{-\alpha}(\beta'+\mu')] \leq 0,
    \qquad
       [\e^{-\alpha}(\mu' - \gamma')] \leq 0,
    \qquad
       [\e^{-\alpha}(\beta'-\gamma') + 2\omega] \leq 0.
\eeq
  Now we can apply these requirements to our configuration at both
  junctions. It turns out that for our purposes it is sufficient to use only
  the third condition which contains the function $\eta = \beta-\gamma$
  given by Eq.\,(\ref{eta}) for solutions with any $V(\phi)$. Using it,
  on $\Sigma_-$ with $x = x_- < 0$ we obtain
\beq                                                          \label{N2-}
       \e^{-\alpha (u_-)} \, \frac{(-s'+ \sign\omega_0)}{s}
       	      + \frac{(1+\Omega_- x)^2}{|x|(1-\Omega_-^2 x^2)} \leq 0,
\eeq
  and on $\Sigma_+$ with $x = x_+ > 0$ we have in a similar way
\beq                                                          \label{N2+}
       \e^{-\alpha(u_+)} \, \frac{(s'- \sign\omega_0)}{s}
       		+ \frac{(1+\Omega_+ x)^2}{x(1-\Omega_+^2 x^2)} \leq 0.
\eeq
  Here $s$ and $s'=ds/du$ refer to the function $s=s(k,u)$ introduced in
  (\ref{def-s}).

  The inequalities (\ref{N2-}) and (\ref{N2+}) lead to the conclusion that
  the matter content of both $\Sigma_+$ and $\Sigma_-$ cannot satisfy the
  NEC (hence also the WEC).

  Indeed, if $\omega_0 > 0$, the inequality (\ref{N2-}) can only hold if
  $1 - s'(k,u) < 0$ at $u=u_-$. But $s'(k,u) = \{\cosh ku, \ 1,\ \cos|k|u\}$
  for $k >0,\ k=0$ and $k < 0$, respectively, and only at $k > 0$ we have
  $1 - s' < 0$. Thus the NEC for $\Sigma_-$ definitely requires $k >0$ in
  the solution valid in the internal region. In a similar way, (\ref{N2+})
  can hold only if $1 - s'(k,u) >0$ at $u=u_+$, and this is only possible if
  $k <0$. All this means that whatever particular solution (with fixed
  parameters including $k$) is taken to describe the internal region, the
  inequalities (\ref{N2-}) and (\ref{N2+}) cannot hold simultaneously.

  If $\omega_0 < 0$, the expression $-s'-1$ appearing in (\ref{N2-}) is
  negative, so this inequality can hold; however, in (\ref{N2+}) there
  instead appears $s'+1 > 0$, hence this inequality cannot hold whatever be
  the parameter $k$.

  We conclude that the NEC is inevitably violated at least on one of the
  surfaces $\Sigma_+$ and $\Sigma_-$.

\section{Conclusion}

  As shown in Ref.\,\citen{BLem09}, static \cyl\ \whs\ with two flat (or
  string) asymptotics in \GR\ can only be obtained with negative matter
  density at least in a certain part of space. We have seen in the present
  study that if we add rotation, there emerge quite a number of new
  phantom-free \wh\ solutions but none of them are \asflat. Even if we try
  to inscribe a rotating \cy\ configuration into flat space by matching it
  to properly chosen parts of flat space, it turns out that this cannot be
  done simultaneously on both entrances to a \wh\ without violating the NEC,
  as has been proven for such a large class of material sources of gravity
  as scalar fields (\ref{L_s}) with any potentials $V(\phi)$ (independently
  of their sign). So the problem of obtaining phantom-free stationary,
  potentially observable \whs\ in \GR\ remains open.

\end{document}